\documentclass[
reprint,
amsmath,
amssymb,
prl,
floatfix]{revtex4-1}

\usepackage{graphicx}
\usepackage{bm}
\usepackage{siunitx}
\usepackage{mathtools}
\usepackage{physics}
\usepackage[allcolors=blue,colorlinks]{hyperref}

\newcommand{\lio}{\mathcal{L}}
\newcommand{\qqq}{\mathcal{Q}}
\renewcommand{\Trace}[2]{ {\rm Tr}_{\rm #1}\left\{ #2 \right \} }

\begin{document}

\title{Pseudospin resonances reveal synthetic spin-orbit interaction}
\author{Christoph Rohrmeier}
\email{christoph.rohrmeier@ur.de}
\author{Andrea Donarini}
\affiliation{Institute of Theoretical Physics, University of Regensburg, 93053 Regensburg, Germany}
\date{\today}

\begin{abstract}
We investigate a spin-full double quantum dot (DQD) coupled to the leads in a pseudospin valve configuration. The interplay of interaction and interference produces in the stability diagram a rich variety of resonances, modulated by the system parameters. In presence of ferromagnetic leads and pseudospin anisotropy, those resonances split, turn into dips and acquire a Fano shape thus revealing a synthetic spin-orbit coupling induced on the DQD. A set of rate equations derived for a minimal model captures those features.
The model accurately matches the numerical results obtained for the full system in the framework of a generalized master equation and calculated within the cotunneling approximation.
\end{abstract}

\maketitle
Quantum dots (QDs) are characterized by a charging energy and by a discrete energy spectrum, both originating from the spatial confinement of their electronic wave-functions. The many-body spectrum of QDs is probed in great detail by coupling them weakly to metallic leads and measuring their transport characteristics as a function of bias and gate voltage. The sequential tunneling (ST) of electrons, hopping from source to drain through the dots, typically produces a differential conductance with Coulomb diamonds decorated by parallel resonant lines which are the spectroscopic signatures of the charging energy and the discrete many-body spectrum. 

Degeneracies \cite{degeneracy} enrich the ST dynamics with interference effects. The latter originate from the coherent superposition of the degenerate states, which arise in this \emph{coherent-sequential-tunneling} (CST) regime and are modulated by the external parameters like the bias and gate voltage. For a spin-full level coupled to non-collinearly polarized ferromagnetic leads (a QD spin valve), interference between the degenerate spin states induces spin accumulation, precession and relaxation, with a resulting non-equilibrium spin polarization of the dot \cite{Koenig2003,Braun2004,Braig2005,Rusinski2005,Weymann2005,Hornberger2008,Zhang2005,Hamaya2007} and spin torque on the leads \cite{Gergs2018}.

For a QD spin valve with almost antiparallel lead polarization, a novel spin resonance has been predicted \cite{Hell2015} within the one-particle Coulomb diamond. A crucial role in this phenomenon is played by the exchange magnetic field \cite{Koenig2003} generated by virtual electronic charge fluctuations between the dot and the leads, i.e.\ the Lamb shift correction to the dot Hamiltonian. Also, orbitally degenerate states support naturally interference if combined with couplings to the leads which mix the tunneling channels \cite{Darau2009,Schultz2010}, as it has been demonstrated for semiconductor wires \cite{Nilsson2010,Karlstroem2011}, QD molecules \cite{Koenig2002, Michaelis2006,Gustavsson2008,Donarini2010,Hatano2011, Niklas2017}, single-molecule junctions \cite{Hettler2003,Begemann2008,Donarini2010,Hettler2003} and suspended carbon nanotubes (CNTs) \cite{Donarini2019}. The control of a QD spin valve is a paradigmatic example of \emph{spintronics}. \emph{Valleytronics} concerns instead the manipulation of a state living in a twofold orbitally degenerate space. Very recently, this concept has been further extended to the one of \emph{flavortronics} \cite{Maurer2020}, for interacting systems with $n$-fold degeneracy.

\begin{figure}
\includegraphics[width=0.75\columnwidth,draft=false]{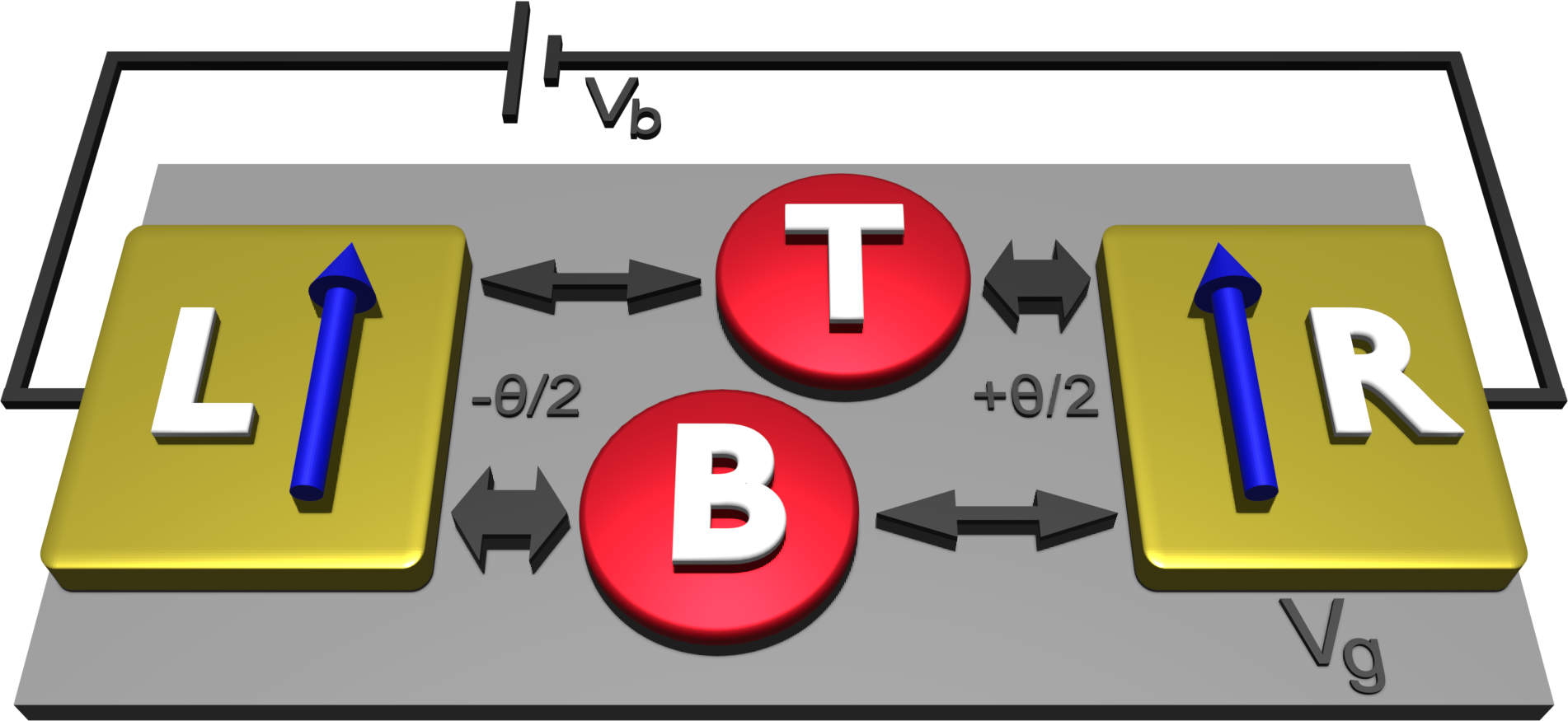} \caption{\label{fig:setup} Schematic setup of a DQD in a pseudospin valve configuration: The left/right lead (L/R) is more strongly coupled to the bottom/top dot (B/T). The angle $\theta\lesssim\pi$ between the pseudospin polarization of the leads ensures the mixing of the pseudospin states. A bias voltage ($V_\text{b}$) applied to the leads and a gate voltage ($V_\text{g}$) control the transport characteristics of the DQD. The blue arrows indicate the parallel spin polarization of the leads.}
\end{figure}

In this Letter, we investigate the interplay between valleytronics and spintronics, between the pseudospin of a DQD with orbital degeneracy and the spin polarization of the ferromagnetic leads. The spatial decay of the Coulomb interaction implies a pseudospin anisotropy on the DQD. In presence of ferromagnetic leads, synthetic spin-orbit interaction emerges. The latter intertwines the spin and the pseudospin degrees of freedom and is revealed by a set of resonances in the stability diagram, which split, turn into dips and acquire a Fano shape by changing the spin polarization of the leads.

\emph{Model} - The spin-full DQD coupled to ferromagnetic leads schematically shown in Fig.~\ref{fig:setup} is described by the system-bath Hamiltonian: $H = H_\text{B} + H_\text{S} + H_\text{T} \label{eq:hamiltonian} $. The bath component reads $H_\text{B}=\sum_{l\sigma k} \varepsilon_{l\sigma k} c_{l\sigma k}^\dagger c_{l\sigma k}$ where $l = {\rm L/R}$ labels the left/right lead, $\sigma$ the spin index and k the momentum both in the lead energy level $\varepsilon_{l\sigma k}$ as well as in the operators $c_{l\sigma k}$. The system Hamiltonian $H_\text{S} = \sum_{r} \left[\left( \si{e} V_\text{g} + \varepsilon^\ast \right)n_{r} +U n_{r} (n_{r}-1) /2\right] +V n_{\text{top}} n_{\text{bot}}$, in which $n_r$ counts the electron number on the top or bottom dot, contains the on-site energy $\varepsilon^\ast$ shifted by a gate voltage $V_\text{g}$ as well as $U$ and $V$, respectively the local and the inter-dot Coulomb interaction. The pseudospin formulation of the system Hamiltonian (cf.\ Supplemental Material) is characterized by a pseudospin anisotropy proportional to $U-V$, essential for the synthetic spin-orbit effects described below. The tunneling Hamiltonian $H_{\text{T}} = \sum_{l\sigma k n} t_{l,n} c_{l\sigma k}^\dagger d_{n} + \text{h.c.}$ combines via the tunneling amplitudes $t_{l,n}$ the bath operators with system operators $d_n$ where $n$ labels a single-particle basis for the DQD.

The CST dynamics of a system with quasi-degenerate many-body spectrum is expressed in terms of tunneling rate matrices \cite{Donarini2019_b,Sigma_tun}. The latter are deduced from $H_\text{T}$ as $\left(\Gamma_l\right)_{n m} = 2 \pi/\hbar \sum_{l\sigma k} t_{l,n}^{\ast}t_{l,m}\delta(\varepsilon-\varepsilon_{l\sigma k})$ and they factorize, in absence of intrinsic spin-orbit coupling, into a spin (s) and an orbital (o) or pseudospin component: \begin{equation}
 \Gamma_l = \Gamma_l^0 \left( \openone_2 +P^\text{s} \bm{n}^{\text{s}}_l \cdot \bm{\sigma}\right) \otimes \left( \openone_2 +P^\text{o} \bm{n}^{\text{o}}_l \cdot \bm{\sigma}\right) 
 \end{equation}
where $\Gamma_l^0$ is the bare tunneling rate for the $l$-lead, $P^\text{s(o)}$ and $\bm{n}^\text{s(o)}_l$ are the strength and the direction vector of the spin (pseudospin) polarization of the lead and $\bm{\sigma}$ is the vector of the Pauli matrices $\sigma_x$, $\sigma_y$ and $\sigma_z$.

We choose parallel spin and almost antiparallel pseudospin directions $\bm{n}^{\text{o}}_\text{L/R}=\left(\cos{\tfrac{\theta}{2}},0,\mp \sin{\tfrac{\theta}{2}}\right)$ with $\theta=0.95 \pi$ (cf.\ Fig.~\ref{fig:setup}). Moreover, we consider high pseudospin polarizations ($P^{\text o} \approx 1$) to achieve an essentially closed pseudospin valve \cite{Hell2015}. Comparable pseudospin polarization strengths have been observed recently in suspended CNTs \cite{Donarini2019}.
 
\emph{Methods} - The transport characteristics are calculated with two complementary approaches. On the one side, a next-to-leading-order expansion in the tunneling coupling is performed using a generalized master equation. To this end, the kinetic equation for the reduced density matrix $\rho_\text{red}=\Trace{B}{\rho}$, i.e.\ the trace over the bath of the total density matrix, is obtained with the Nakajima-Zwanzig projector technique \cite{Nakajima1958,Zwanzig1960}. The steady state is defined by $\dot{\rho}^{\infty}_{\text{red}} = 0 = \left( \lio_{\text{S}}+\mathcal{K} \right) \rho^{\infty}_{\text{red}}$ where the Liouville superoperator, in general defined as $\mathcal{L} \rho = -\tfrac{i}{\hbar}\comm{H}{\rho}$, is taken here with respect of $H_\text{S}$. The Kernel superoperator $\mathcal{K}$ reads
\begin{equation}\label{eq:kernel}
\mathcal{K} \rho^{\infty}_{\text{red}} = \text{Tr}_{\text{B}} \left\lbrace \lio_{\text{T}} \sum_{n=0}^{\infty} \left(\tilde{\mathcal{G}}_0 \qqq \lio_{\text{T}} \qqq \right)^{2n} \tilde{\mathcal{G}}_0 \lio_{\text{T}} \rho^{\infty}_{\text{red}}\otimes \rho_{\text{B}}\right\rbrace
\end{equation}
where $\qqq = 1 - \mathcal{P}$ with $\mathcal{P} = \Trace{B}{\bullet} \otimes \rho_{\rm B}$ is a Nakajima-Zwanzig projector, $\rho_{\rm B}$ is the equilibrium density operator for the bath and $\tilde{\mathcal{G}}_0 $ is the Liouville space propagator in absence of tunneling coupling \cite{Koller2010_1,Leijnse2010,Koller2010_2,Niklas2018}. The first term of the sum in \eqref{eq:kernel} reproduces the ST regime. We consider here a truncation up to the cotunneling regime ($n = 1$; cf.\ Supplemental Material). Eventually, from the stationary density matrix $\rho_{\rm red}^\infty$, we calculate the stationary current at lead $l$ as $I_l = \Trace{\rm S}{\mathcal{K}_l\rho_{\rm red}^\infty}$ where the current Kernel is obtained from the propagator kernel in \eqref{eq:kernel} by changing the leftmost tunneling Liouvillean with the current operator \cite{Koller2010_1,Leijnse2010,Koller2010_2,Niklas2018}. A novel treatment of the cotunneling integrals founded on the work of \cite{Koller2010_2,Koller2010_1,Mantelli2016,Milgram2017} allowed us for the implementation of a transport code which includes all coherences necessary to capture the interference effects in our system. Moreover, a systematic test of robustness for such effects beyond the ST approximation has been achieved.

In a complementary approach, we set up a minimal model in the regime of CST (cf.\ \cite{Maurer2020}). As we focus on the resonance between zero and one particle, we restrict here to the coupled dynamics of the populations $p_0$ and $p_\sigma$ (empty and singly occupied DQD with spin $\sigma$) complemented by one of the pseudospin vectors $\bm{T}_\sigma$:
\begin{eqnarray}
	 \dot p_0 &=&~ -4 \gamma^+ p_0+\sum_{\sigma} D_\sigma \left[ \gamma^- p_{\sigma}+2 \bm{\gamma}^- \cdot \bm{T}_{\sigma} \right], \label{eq:p0} \\ 
	 \dot p_\sigma &=&~ D_\sigma \left[ 2 \gamma^+ p_0-\gamma^- p_{\sigma}-2 \bm{\gamma}^- \cdot \bm{T}_{\sigma}\right], \label{eq:psigma}\\
	 \dot{\bm{T}}_{\sigma} &=&~ D_\sigma \left[-\gamma^- \bm{T}_{\sigma} + \bm{\gamma}^+ p_0 -\tfrac{1}{2} \bm{\gamma}^- p_\sigma \right]+\bm{B}_\sigma \times \bm{T}_{\sigma} \label{eq:pspinpart}
\end{eqnarray}
where $\bm{\gamma}^\pm=\sum_l \Gamma_l^0 f^\pm_l(\varepsilon) \bm{n}_l^\text{o}$, $\gamma^\pm=\sum_l \Gamma_l^0 f^\pm_l(\varepsilon)$ and $D_{\uparrow(\downarrow)}=1\pm P^\text{s}$. The Fermi-functions are dependent on the temperature $T$ with $k_\text{B}$ the Boltzmann constant, the chemical potential $\mu_l$ of the lead $l$ and $p=\pm$ which indicates in- or out-tunneling: $f_l^{p} \left( \varepsilon \right) = 1 /(e^{p(\varepsilon-\mu_l)/(k_\text{B}T)}+1)$. The term $2 \bm{\gamma}^- \cdot \bm{T}_{\sigma}$ in \eqref{eq:p0}-\eqref{eq:psigma} ensures the coupling of the populations and the accumulated pseudospin. Three conceptually different mechanisms yield the time evolution of the pseudospin: the first term in \eqref{eq:pspinpart} describes relaxation, accumulation due to changes in the populations characterizes the following two terms. The last term contains the \emph{spin dependent} pseudo exchange field $\bm{B}_\sigma$ which, analogously to magnetic fields, generates pseudospin precession. The exchange field is defined as 
	\begin{eqnarray}
	\bm{B}_{\sigma}&&= \sum_l 2 P^\text{o} \Gamma_l^0 \left[D_\sigma \left(p_l\!\left(E_1-E_0\right)-p_l\!\left(E_\text{2g}-E_1\right)\right) \bm{n}_l^{\text{o}} \right. \nonumber \\
	&&\left. + D_{\bar \sigma} \left(p_l\!\left(E_\text{2e}-E_1\right)-p_l\!\left(E_\text{2g}-E_1\right)\right) \left( \bm{n}_l^{\text{o}} \cdot \bm{e}_z\right)\bm{e}_z \right]\label{eq:exchangefield}
	\end{eqnarray}
 with $p_l\!\left(x\right)=\text{Re} \Psi^{(0)}( \tfrac{1}{2}+\tfrac{i \left(\si{e}V_\text{g}+ x- \mu_l\right)}{2 \pi k_\text{B}T})$ where $\Psi^{(0)}(z)$ is the digamma-function. The subscript of the energy $E_x$ labels the one-particle state (1) and the two-particle excited/ground state ($\text{2e}/\text{2g}$). It is crucial to include in the exchange field the two-particle energies containing $U$ and $V$, even though we do not account for the populations of those states. Also, energy levels far from the ST resonance ($\Delta E/k_B T \gg 1$) do influence the exchange field due to the logarithmic tails of the digamma-functions.

\begin{figure*}
 \includegraphics[width=474pt]{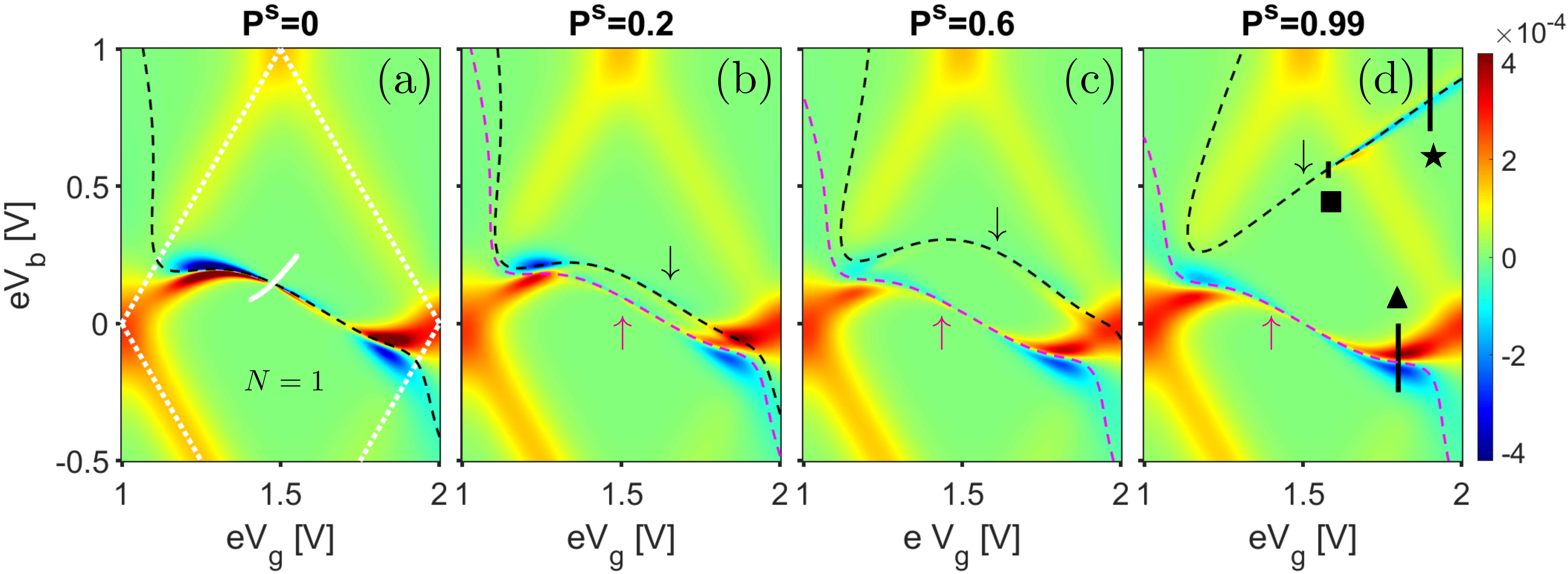}
\caption{\label{fig:pspins} Differential conductance shows pseudospin resonances in a DQD and is tuned by spin polarization $P^s$: The one-particle diamond is highlighted by the dotted white lines in panel (a). The three vertical black lines ($\bigstar,\blacksquare,\blacktriangle$) indicate the bias traces of Fig.~\ref{fig:biascuts}. The dashed magenta (black) line is the resonance condition of the $\uparrow$($\downarrow$)-electrons (cf.\ \eqref{eq:resonancecondition}). The solid white line indicates the minimum of $B_{\sigma,\perp}$ which matches perfectly a local minimum within the pseudospin resonance. The parameters are the following: $U=2 V$, $k_{\text{B}}T=0.05 V$, $P^o=0.99$, $\theta=0.95 \pi$, $\Gamma_R=2.5 \times 10^{-3} V=2 \Gamma_L$, $\varepsilon^\ast=-2 V$ and $ W= 250 V $.}
\end{figure*}

\emph{Results} - In Fig.~\ref{fig:pspins} stability diagrams of a DQD in the cotunneling regime are displayed for several spin polarizations of the leads. We focus on the one-particle Coulomb diamond, highlighted in panel (a) by the dotted white lines. Here we would normally expect an essentially fixed particle number and, due to Coulomb repulsion, only an exponentially suppressed current. An exception to this rule can be clearly seen in panel (a) where a distinctive resonance, highlighted by the dashed black line, is cutting through the Coulomb diamond. Increasing the spin polarization $P^\text{s}$ (Fig.~\ref{fig:pspins}\,(b)-(d)) leads to a splitting of this resonance, marked by the dashed lines. In the upper right corner of Fig.~\ref{fig:pspins}\,(d), a resonance can be observed even outside the diamond. 
This transport effect is explained by pseudospin resonances in analogy to the spin resonances reported in \cite{Hell2015}. The pseudospin is associated with the orbital degree of freedom of the DQD. In our setup, the orbital polarizations of the leads are almost antiparallel thus resulting in an almost closed pseudospin valve. The latter is indicated in Fig.~\ref{fig:setup} by the different sizes of the arrows connecting the leads and the dots. Solely varying the coupling strength would correspond to a sweep of the lead polarization along the $z$-direction. Pseudospin resonances require, instead, non-collinear orbital polarizations as well as an asymmetry in the bare coupling strength $\Gamma_l^0$ between the right and left lead. The latter shifts the resonance away from the zero bias line \cite{Hell2015}. The necessary $\sigma_x$ or $\sigma_y$ orbital polarization of the leads translates into non-diagonal $\Gamma_l$-matrices, which can be interpreted as tunneling to a coherent superposition of two different orbitals. Experimental evidence of such coherent superpositions for QDs in the weak tunneling regime has been reported \cite{Nilsson2010,Donarini2019}. 
In the framework of \eqref{eq:p0}-\eqref{eq:pspinpart}, vectorial resonance conditions can be formulated similarly to the ansatz in \cite{Hell2015,Maurer2020}: 
\begin{equation}
 \bm{B}_\sigma \cdot \left(\bm{n}_\text{L}^\text{o}-\bm{n}_\text{R}^\text{o}
 \right)=0\label{eq:resonancecondition}.
\end{equation}
The spin dependent exchange field generates two distinct conditions, each determining the position of the corresponding resonance in the $V_\text{g}$-$V_\text{b}$-plane: the magenta (black) dashed line in Fig.~\ref{fig:pspins} for the $\uparrow(\downarrow)$-electrons. The accuracy of \eqref{eq:resonancecondition} in determining the resonance positions reduces as $\theta$ is chosen further away from antiparallel alignment. In contrast to the resonance conditions formulated in \cite{Hell2015} and in \cite{Maurer2020}, we choose \eqref{eq:resonancecondition}, where the drain and the source equally participate, since it matches the numerical resonances on a broader parameter range.
Despite the subtle differences, though, all three conditions mentioned above can only predict the \emph{position} of the resonances, but not their \emph{character}. The same resonance condition corresponds to a dip in the current ($\bigstar$ in Fig.~\ref{fig:pspins}), or to a peak ($\blacktriangle$) and even to a Fano-like asymmetric peak-dip ($\blacksquare$). Finally, the current peak is strongly modulated along the same resonance line and it can even disappear, as exemplary highlighted in panel (a) of Fig.~\ref{fig:pspins} with the solid white line. The discovery and explanation of such \emph{qualitative} differences in the pseudospin resonances, which originate from the intertwining of spin and pseudospin, represent the main result presented in this Letter.

\begin{figure}
\includegraphics[width=4cm,draft=false]{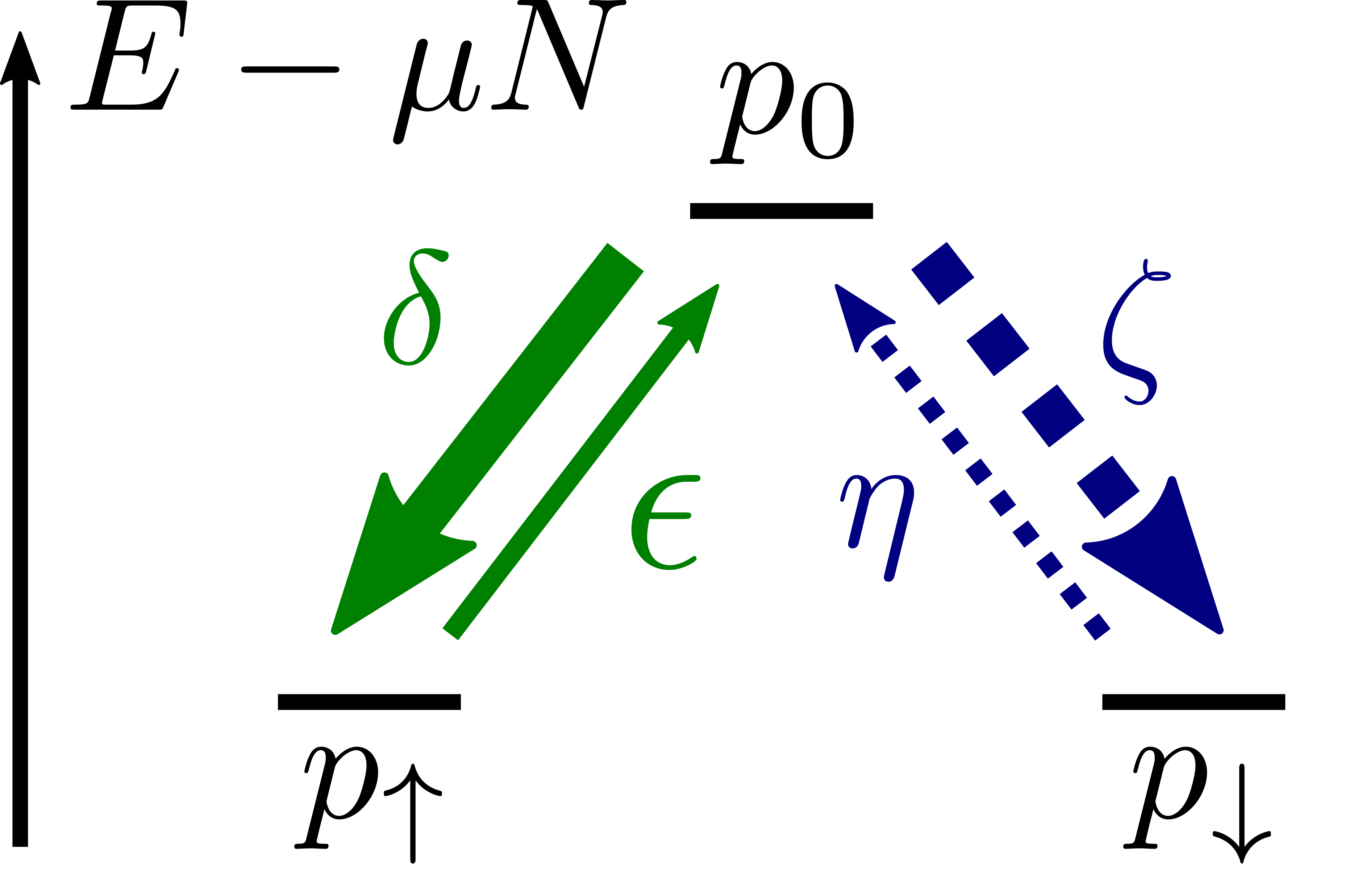} \caption{\label{fig:ratescheme} Rate scheme of the three populations $p_0$, $p_\uparrow$ and $p_\downarrow$: The four arrows indicate the rates between the populations while their size specifies the strength of them. The dashed rates for the minority spin are furthermore lowered by the majority spin polarization of the leads.}
\end{figure}

\begin{figure}[hbt!]
 \includegraphics[width=0.82\columnwidth,draft=false]{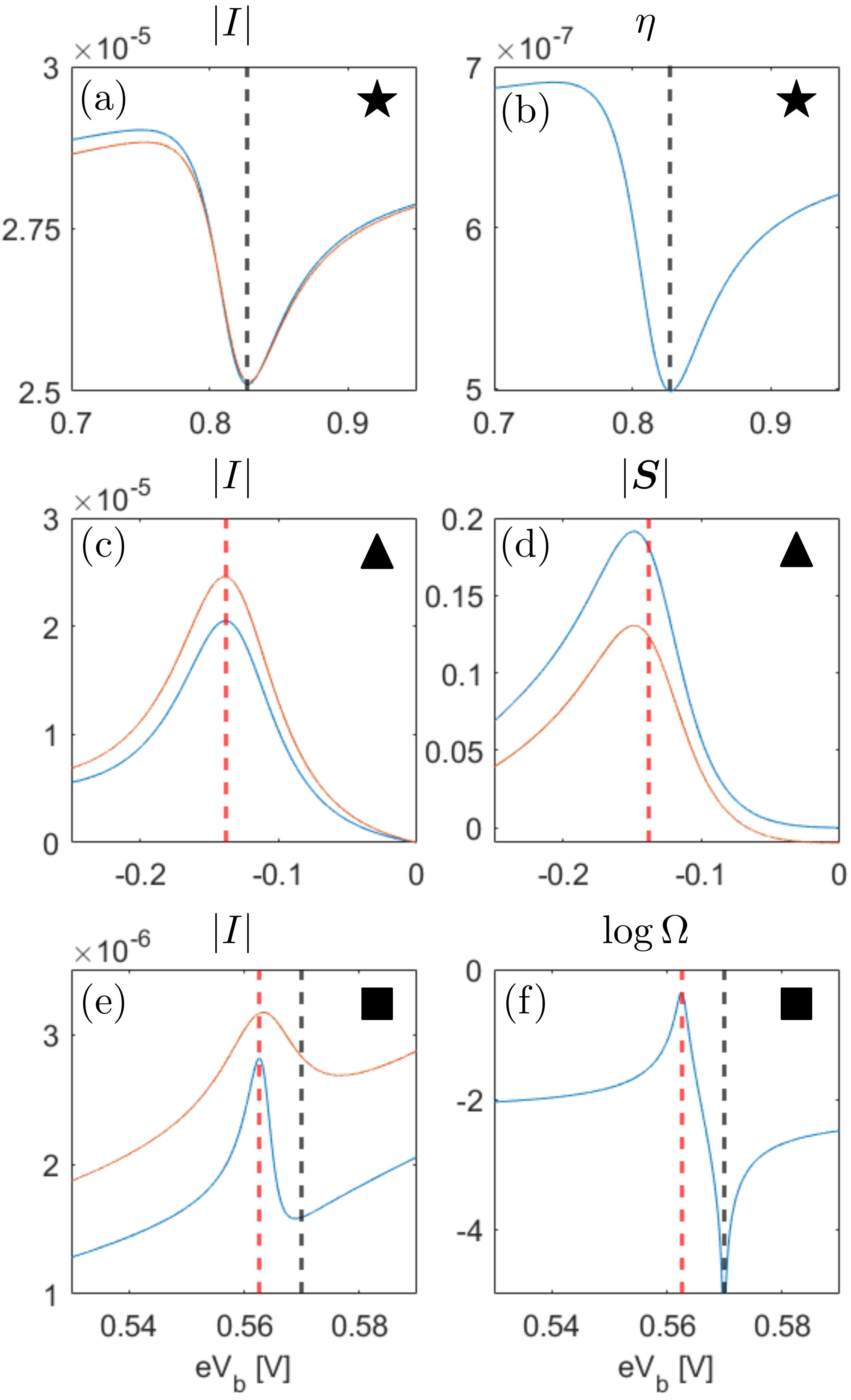}
\caption{\label{fig:biascuts} Effective rate analysis of the bias traces from Fig.~\ref{fig:pspins}\,(d): The absolute value of the current shows (a) a dip at $\si{e}V_\text{g}=1.9 V$, (c) a peak at $\si{e}V_\text{g}=1.8 V$ and (e) a Fano-like shape at $\si{e}V_\text{g}=1.58 V$. The analytic solution of the effective ST model is depicted in blue whereas the orange line shows the full cotunneling calculations. The black (red) dashed lines indicate the position of the minimum of $|B_{\sigma,\perp}|$ ($|B_{\sigma,\parallel}|$) and correspond to a minimum (maximum) of the current. (b) The rate $\eta$ strongly correlates to the current. (d) The absolute value of the spin of our system $|\bm{S}|$ is following the trend of the current. (f) The logarithm of the ratio $\Omega = B^2_{\downarrow,\perp}/\left(a_\downarrow^2+B^2_{\downarrow,\parallel}\right)$ highlights the two extrema of $\Omega$ which result in a peak and a dip in the current.}
\end{figure}

For a deeper understanding of the numerical data of Fig.~\ref{fig:pspins}, we further elaborate on the equations of motion of \eqref{eq:p0}-\eqref{eq:pspinpart}.
Solving \eqref{eq:pspinpart} in the stationary limit leads to $\bm{T}_\sigma=\tfrac{a_\sigma}{a_\sigma^2+\bm{B}_\sigma^2}\left(\bm{b}_\sigma+\tfrac{\bm{B}_\sigma \cdot \bm{b}_\sigma}{a_\sigma^2}\bm{B}_\sigma + \tfrac{\bm{B}_\sigma \times \bm{b}_\sigma}{a_\sigma}\right)$ with $a_\sigma=D_\sigma \gamma^-$ and $\bm{b}=D_\sigma\left(\bm{\gamma}^+ p_0 -\tfrac{1}{2} \bm{\gamma}^- p_\sigma \right)$.
By substituting $\bm{T}_\sigma$ into \eqref{eq:p0}-\eqref{eq:psigma}, the problem is reduced to a set of effective rate equations for the populations $p_0$ and $p_\sigma$ with the transition rates schematically indicated in Fig.~\ref{fig:ratescheme}. The stationary current reads, correspondingly, 
\begin{equation}
\label{eq:current}
	I_\text{L} =4 \gamma^+_L p_0+\sum_{\sigma} D_\sigma \left( -\gamma^-_\text{L} p_{\sigma}-2 \bm{\gamma}^-_\text{L} \cdot \bm{T}_{\sigma} \right).
\end{equation}
The panels (a), (c) and (e) of Fig.~\ref{fig:biascuts} show a direct comparison between the absolute value of the current as obtained form the full numerical calculation (orange) and the analytical approach (blue) of \eqref{eq:current}.
In all three cases, the analytical result well reproduces the qualitative behavior of the current and the position of its extrema.

In a simple physical picture, we expect a peak in the current whenever the pseudospin precession caused by the exchange field releases the blockade induced by the pseudospin valve. A dip arises, instead, whenever this mechanism is locally suppressed. Both phenomena happen in close vicinity to the aforementioned resonance condition \eqref{eq:resonancecondition}. Only the analysis of the effective rates represented in Fig. \ref{fig:ratescheme} allows, though, to distinguish them. 
The incoherent superposition of a minority and majority spin channel yields the current. Its modulation is determined by the depopulation rates $\epsilon$ and $\eta$. Thus, as confirmed by the resemblance between panels (a) and (b) in Fig.~\ref{fig:biascuts}, the shape of a $\downarrow$-resonance, is given by the \emph{bottleneck} rate
\begin{equation}
\eta = D_\downarrow \gamma^- \left[1- \frac{|\bm{\gamma}^-|^2}{(\gamma^-)^2} \frac{1}{1+\frac{B^2_{\downarrow,\perp}}{a_\downarrow^2+B^2_{\downarrow,\parallel}}} \right] \label{eq:etarate}
\end{equation}
with $B_{\downarrow,\parallel}^2=\left(\bm{B}_{\downarrow} \cdot \bm{\gamma}^-\right)^2/ |\bm{\gamma}^-|^2$ and $B^2_{\downarrow,\perp}=\bm{B}_{\downarrow}^2 -B^2_{\downarrow,\parallel} $ the exchange field components parallel and perpendicular to $\bm{\gamma}^-$. In itself, $\eta$ is strongly influenced by the ratio $\Omega = B^2_{\downarrow,\perp}/(a_\downarrow^2+B^2_{\downarrow,\parallel})$ in which the proposed physical explanation based on the precession dynamics is encoded. In absence of the perpendicular pseudo magnetic field component, no precession occurs and the bare pseudospin valve factor $|\bm{\gamma}^-|^2/(\gamma^-)^2$ reduces the rate. The other extreme is reached when the ratio $\Omega$ peaks, therefore suppressing the pseuodspin valve factor. Such phenomenon only occurs if the absolute value of the parallel component $|B_{\downarrow,\parallel}|$ is minimized, since the dephasing rate $a_\downarrow$ is proportional to a Fermi-function, which varies smoothly within the Coulomb diamond.

The dashed lines in Fig.~\ref{fig:biascuts} show the accuracy of the precession argument in determining the position of the current extrema. The rate $\epsilon$, obtained by replacing $\downarrow$ with $\uparrow$ in \eqref{eq:etarate}, is used for the panels (c) and (d) of Fig.~\ref{fig:biascuts}. In Fig.~\ref{fig:biascuts}\,(e), both the suppression and the enhancing of the current appear in close vicinity and form a Fano-like line shape. In order to emphasize the rather weak dip, we depicted in Fig.~\ref{fig:biascuts}\,(f) the logarithm of the ratio $\Omega$. The ratio $\Omega$ has two extrema which stem from minima of the corresponding exchange field components $|B_{\downarrow,\perp}|$ and $|B_{\downarrow,\parallel}|$. Despite its superficial resemblance to a Fano resonance, the origin of this peak-dip current resonances cannot be ascribed to the interference processes typical of Fano resonances, also seen in QD setups \cite{Baranski2020,Torio2004,Baernthaler2010,Joe2005,Johnson2004}. 

Moreover, the relevance of $\Omega$ decreases if $a_\sigma \gg |\bm{B}_\sigma|$, i.e.\ when the dephasing rate exceeds the precession frequency and the direction of the exchange field becomes irrelevant for the transport. Thus, no resonances appear on the left upper corner in correspondence to the black and magenta dashed lines of the panels (a)-(d) of Fig.~\ref{fig:pspins} even if they would be predicted by the resonance condition \eqref{eq:resonancecondition}. 

\emph{Conclusion} - A DQD weakly coupled to ferromagnetic leads in pseudospin valve configuration is characterized by a rich variety of pseudospin resonances. They decorate the Coulomb diamonds with novel features which range from a peak to a dip to a Fano shape in the current. These transport characteristics reveal the synthetic spin-orbit interaction induced on the system by the interplay of leads polarization and pseudospin anisotropy on the DQD. 

The cotunneling calculations ensure the robustness of such effect beyond the CST limit. Moreover, with the help of a minimal model, we give an accurate physical picture of the resonances and relate their position and character to a precession dynamics which modulates the pseudospin valve effect. The generality of the model allows for its applicability to the wide class of nanoscale junctions with orbital degeneracy, including e.g.\ single-molecules junctions or CNT-QDs. Particularly, coherent population trapping and signatures of pseudospin precession have been recently demonstrated in a CNT with a tunneling coupling similar to the one proposed here \cite{Donarini2019}.

\emph{Acknowledgments} - The authors acknowledge financial support from the Elite Netzwerk Bayern via the IGK Topological Insulators and the Deutsche Forschungsgemeinschaft via the SFB 1277 (subprojects B02 and B04). We thank moreover M.\ Grifoni for fruitful discussions.

\bibliography{bibliography}

\begin{thebibliography}{40}
\bibitem [{deg()}]{degeneracy}%
  \BibitemOpen
  \href@noop {} {\enquote {\bibinfo {title} {Effectively degenerate are all
  states with an energy separation smaller than the tunnelling induced
  broadening.}}\ }\BibitemShut {NoStop}%
\bibitem [{Sig()}]{Sigma_tun}%
  \BibitemOpen
  \href@noop {} {\enquote {\bibinfo {title} {In another context, these matrices
  are called tunnelling self-energies.}}\ }\BibitemShut {NoStop}%
\end{thebibliography}
\end{document}


\title{Supplemental Material: Pseudospin resonances reveal synthetic spin-orbit interaction}

\author{Christoph Rohrmeier}
\email{christoph.rohrmeier@ur.de}
\author{Andrea Donarini}%
\affiliation{Institute of Theoretical Physics, University of Regensburg, 93053 Regensburg, Germany}
\date{\today}

\maketitle
\section{Pseudospin Anisotropy}
The interference effects presented in the main text are inherently tied to the anisotropy of the pseudospin in the double quantum dot. In this section, we reformulate the system Hamiltonian to highlight such anisotropy. The system Hamiltonian is defined as $H_\text{S} = \sum_{r} \left[\left( \si{e} V_\text{g} + \varepsilon^\ast \right)n_{r}  +U n_{r} (n_{r}-1) /2 \right]+ V n_{\text{top}} n_{\text{bot}}$ where $n_r=\sum_\sigma d_{r\sigma}^\dagger d_{r\sigma}$ represents the number operator for the $r$-dot with $d_{r\sigma}$ the annihilation operator for an electron on the top or bottom dot with spin $\sigma$, $U$ the local Coulomb repulsion, $V$ the inter-site Coulomb repulsion, $\varepsilon^\ast$ the on-site energy, $V_\text{g}$ the gate voltage and $\si{e}$ the elementary charge. 
The collective index $n$ of the system is here explicitly written in terms of its orbital and spin  components $r$ and $\sigma$.
We want to express this Hamiltonian in terms of the pseudospin of the system. We use therefore the following definitions for the total number operator $N=\sum_{r}n_r$ and the $z$-component for the pseudospin $T_z=\sum_{\sigma r r'} \tfrac{1}{2} d^\dagger_{r'\sigma} (\sigma_z)_{rr'}  d_{r\sigma}$ where $\sigma_z$ is the $z$-Pauli matrix:
\begin{align}
    n_\text{top}&=\frac{n_\text{top}+n_\text{bot}}{2}+\frac{n_\text{top}-n_\text{bot}}{2}\coloneqq \frac{N}{2}+T_z,\\
    n_\text{bot}&=\frac{n_\text{top}+n_\text{bot}}{2}-\frac{n_\text{top}-n_\text{bot}}{2}\coloneqq \frac{N}{2}-T_z
\end{align}
 to obtain
\begin{equation}
    H_\text{S} = \left( \bar \varepsilon -\frac{U}{2}\right)N+\frac{U+V}{4}N^2+\left(U-V\right)T_z^2\label{eq:systemhamiltonian}
\end{equation}
where $\bar \varepsilon=\si{e} V_\text{g} + \varepsilon^\ast$.
In this representation of the Hamiltonian, it is evident that the difference of the local and inter-site Coulomb repulsion translates into an easy-plane anisotropy of the pseudospin, i.e.\ it is energetically more favorable for the pseudospin vector to point in the $\sigma_x$-$\sigma_y$-plane than to point in the $\sigma_z$-direction where one has to pay extra energy to localize the electrons on one dot. Interestingly, a top-bottom tunneling $t$ in the Hamiltonian could be seen in the framework of pseudospin as a pseudo-magnetic field:
\begin{equation}
    H_\text{S} = \left( \bar \varepsilon -\frac{U}{2}\right)N+\frac{U+V}{4}N^2+\left(U-V\right)T_z^2+\bm{B}_{\rm t}\cdot\bm{T}
\end{equation}
with $B_{{\rm t},x}=2\text{Re}\;t$, $B_{{\rm t},y}=2\text{Im}\;t$ and $B_{{\rm t},z}=0$ associated to a top-bottom tunnelling amplitude $t$. This tunnelling process would lift the orbital degeneracy of our system and thus destroy the pseudospin resonances if the magnitude of such Zeeman-like splitting is big enough. We argue that a small hopping $|t|<\hbar \Gamma$ is not detrimental as the coupling to the leads $\Gamma$ can not resolve the lifted degeneracies. Therefore, one still expects interference effects to appear and the pseudo-magnetic field $\bm{B}_{\rm t}$ would simply add to the exchange one generated by the tunnelling to the leads. 

\section{Lamb shift Hamiltonian}
\begin{figure*}
 \includegraphics[width=280pt]{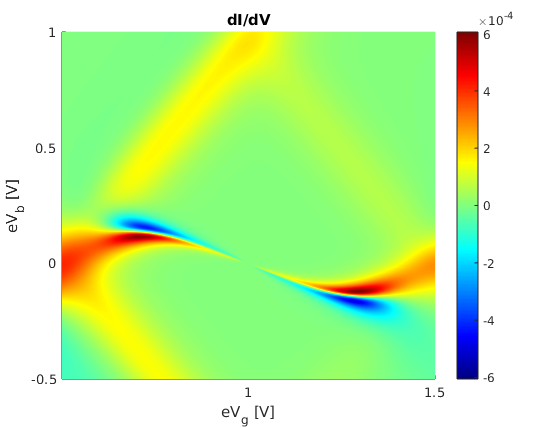}
\caption{\label{fig:dIdVwithoutpspinanisotropy} Differential conductance for $P^s=0.99$ and constant interaction ($U=V$): The stability diagram of a double quantum dot where $U=V=1$ shows only one pseudospin resonance in comparison with Fig.~2\,(d) of the main text where two resonances for the different spin species appear. The plot should highlight the fact that the pseudospin anisotropy is the main cause for the synthetic spin-orbit effect. Changing the spin polarisation at constant interaction does not alter the differential conductance in agreement with (7) of the main text. The calculation is performed with the same cotunneling code as in Fig.~2 of the main text. The chosen parameters are the following: $U=1 V$, $k_{\text{B}}T=0.05 V$, $P^o=0.99$, $\theta=0.95 \pi$, $\Gamma_R=2.5 \times 10^{-3} V=2 \Gamma_L$, $\varepsilon^\ast=-2 V$ and $ W= 250 V $.}
\end{figure*}

In this section, we deduce the Lamb shift Hamiltonian for the one-particle subspace with the help of the reformulated system Hamiltonian. The $\alpha$-component of the spin $\bm{S}$ and the pseudospin $\bm{T}$ operators in the one-particle subspace are defined by
\begin{align}
    \left(\mathcal{P}_1 S_\alpha \mathcal{P}_1\right)_{r\sigma, r'\sigma'}&=\frac{1}{2} \delta_{r r'} \left(\sigma_{\alpha}\right)_{\sigma\sigma'}, \\
     \left(\mathcal{P}_1 T_\alpha \mathcal{P}_1\right)_{r\sigma, r'\sigma'}&=\frac{1}{2} \delta_{\sigma\sigma'} \left(\sigma_{\alpha}\right)_{r r'}
\end{align}
where $\sigma_\alpha$ are the Pauli matrices and $\mathcal{P}_x=\sum_{\ell \sigma}\ket{\ell \sigma}\bra{\ell \sigma}$ are the projector operators of the $x$-particle subspace. For example $\mathcal{P}_1$ runs over the four states $\ket{\text{top} \uparrow}$,  $\ket{\text{bot} \uparrow }$, $\ket{\text{top} \downarrow }$ and $\ket{\text{bot} \downarrow }$.
The principal values stemming from the energy integration are denoted in the following by $p_l\!\left(x\right)=\text{Re} \Psi^{(0)}( \tfrac{1}{2}+\tfrac{i \left(\si{e}V_\text{g}+  x- \mu_l\right)}{2 \pi k_\text{B}T})$ where $\Psi^{(0)}(z)$ is the digamma-function. The Lamb shift Hamiltonian reads then
\begin{equation}
   H_\text{LS}=\sum_{l r \sigma r' \sigma'} \left(\Gamma_{l}\right)_{r\sigma,r'\sigma'} \mathcal{P}_1\left[d_{r\sigma}^\dagger p_l\left(E_1-H_S\right)d_{r' \sigma'}+d_{r' \sigma'} p_l\left(H_S-E_1\right)d^\dagger_{r \sigma}\right]\mathcal{P}_1 .
\end{equation}
We insert the system Hamiltonian (cf.\ \eqref{eq:systemhamiltonian}), and perform a Taylor expansion with respect to the anisotropy component. Exploiting the relation $\mathcal{P}_2 T_z^2 \mathcal{P}_2=\left(\mathcal{P}_2 T_z^2 \mathcal{P}_2\right)^n$ for $n\geq1$, we can simplify the terms containing $ T_z^2$:
\begin{align}
    \mathcal{P}_2 p_l\left(\bar \varepsilon + V -\mu_l+ (U-V)T_z^2\right)\mathcal{P}_2&=\mathcal{P}_2 p_l\left(\bar \varepsilon + V -\mu_l\right)+\sum_{n=1}^\infty\frac{1}{n!}p_l^{(n)}\left(\bar \varepsilon + V -\mu_l\right)  \left(U-V\right)^n  \mathcal{P}_2 T_z^2   \mathcal{P}_2 \nonumber \\
    &=\mathcal{P}_2 \left[p_l\left(\bar \varepsilon + V -\mu_l\right)+T_z^2 \left( p_l\left(\bar \varepsilon + U -\mu_l\right)-p_l\left(\bar \varepsilon + V -\mu_l\right)\right)   \right]\mathcal{P}_2 \nonumber \\
     &=\mathcal{P}_2 \left[p_l\left(E_\text{2g}-E_\text{1}\right)+T_z^2 \left( p_l\left(E_\text{2e}-E_\text{1}\right)-p_l\left(E_\text{2g}-E_\text{1}\right)\right)   \right]\mathcal{P}_2. 
\end{align}    
We further simplify the term containing $T_z^2$ using the relation
\begin{equation}
    \mathcal{P}_1 d_{r'\sigma'} T_z^2 d^\dagger_{r\sigma}\mathcal{P}_1=\frac{1}{2}\mathcal{P}_1 d_{r'\sigma'}d_{r\sigma}^\dagger \mathcal{P}_1  + \sum_k (\sigma_z)_{k r} \mathcal{P}_1 d_{r'\sigma'}d_{k\sigma}^\dagger  T_z \mathcal{P}_1.
\end{equation}
Some algebra leads eventually to the following formulation of the Lamb shift Hamiltonian, obtained under the additional assumption of parallel spin polarisation of the leads: 
\begin{align}
   H_\text{LS}=&\sum_l  \Gamma_l^0 \left[ p_l\left(E_1-E_0\right)+2p_l\left(E_{\text{2g}}-E_1\right)+p_l\left(E_{\text{2e}}-E_1\right)\right] \mathcal{P}_1 \nonumber\\
   &+\sum_l \Gamma_l^0 \left[ p_l\left(E_1-E_0\right)-p_l\left(E_{\text{2g}}-E_1\right)\right] P^\text{s} \bm{n}^\text{s}_l \cdot \mathcal{P}_1 \bm{S} \mathcal{P}_1 \nonumber\\
   &+\sum_l  (1+P^\text{s})  \Gamma_l^0  \left(p_l\left(E_1-E_0\right)-p_l\left(E_{\text{2g}}-E_1\right)\right)P^\text{o}_l \bm{n}^\text{o}_l \cdot \mathcal{P}_1 \bm{T} \mathcal{P}_1 \nonumber \\
   &+\sum_l (1-P^\text{s}) \Gamma_l^0  \left(p_l\left(E_\text{2e}-E_1\right)-p_l\left(E_{\text{2g}}-E_1\right)\right)P^\text{o}_l \left(\bm{n}^\text{o}_{l}\right)_z  \mathcal{P}_1 T_z \mathcal{P}_1 .\label{eq:LS}
\end{align}
The first two terms do not contribute to the time evolution of the reduced density matrix in the one-particle subspace, as the parallel spin polarization defines a common quantization axis. Finally, we transform the Lamb shift induced dynamics for the reduced density matrix into equations of motion for the pseudospins: 
\begin{equation}\left(\dot \rho_\text{red,1}^\infty\right)_\text{LS} = \frac{i}{2 \pi} \comm{H_\text{LS}}{\rho_\text{red,1}^\infty} \Longleftrightarrow \left(\bm{\dot T}_\sigma \right)_\text{LS}=\bm{B}_\sigma \cross \bm{T}_\sigma.
\end{equation}
Assuming constant interaction ($U=V$), the energies $E_\text{2e}$ and $E_\text{2g}$ coincide. This implies that also the last term of \eqref{eq:LS} vanishes. Consequently, only one pseudo spin resonance is present  in the stability diagram (cf.\ Fig.~\ref{fig:dIdVwithoutpspinanisotropy}).

\section{Transport theory}
We outline in this section the transport theory used for the calculation of the cotunneling current presented in Fig.~2 and Fig.~4 of the main text. The starting point of the derivation is the Liouville-von Neumann equation which, written in terms of the Liouville superoperator $\mathcal{L}(t)$, reads:
\begin{equation}\label{lvne2}
\mathcal{L}(t) X \coloneqq  -\frac{i}{\hbar}\comm{H(t)}{X} \implies \dot{\rho}(t) = \mathcal{L}(t) \rho(t)
\end{equation}
where $H(t)$ is the - in general - time-dependant Hamiltonian and $\dot \rho(t)$ is time-derivative of our density matrix. The Liouville superoperator acts on a generic matrix $X$ in analogy to a matrix which acts on a vector. In the case of the time-independent Hamiltonian  $H$ which we consider in this Letter, the Liouvillian can be written as  $\mathcal{L}= \mathcal{L}_{\text{B}}+\mathcal{L}_{\text{S}}+\mathcal{L}_{\text{T}}$. Based on \eqref{lvne2}, we can deduce a generalized master equation (GME) which can accurately describe the time evolution of our system while rigorously taking into account not only the populations of the underlying density matrix but also all its coherences. Only then it is possible to capture interference effects in all their facets. The method of choice to derive a GME is the Nakajima-Zwanzig projection operator technique \cite{Nakajima1958,Zwanzig1960}.
The main idea of Nakajima and Zwanzig is to split the total density into two parts: one where the quantum dot system and the leads are separated ($\mathcal{P}\rho_{\text{tot}}$) and one where the entanglement of the quantum dot system and the leads is captured ($\mathcal{Q}\rho_{\text{tot}}$). This is done by the two projectors
\begin{align}\label{nakajimizwanzigdef}
\mathcal{P} X &\coloneqq \text{Tr}_{\text{B}} \left\lbrace X \right\rbrace \otimes  \rho_{\text{B}}, \\
\mathcal{Q} X &\coloneqq  (1-\mathcal{P}) X
\end{align}
with $X$ an arbitrary density matrix and $\text{Tr}_{\text{B}}$ is the trace over the bath. 
The Nakajima-Zwanzig equation reads
\begin{equation} \label{eq:nakajimazwanzigeq}
\mathcal{P}\dot{\rho}(t) = \mathcal{L}_{\text{S}}\mathcal{P}\rho(t)+ \int_{0}^{t} \diff s \; \mathcal{K}(t-s) \pp \rho(s)
\end{equation}
with the Kernel superoperator $\mathcal{K}(t)= \pp \lio_{\text{T}} \bar{G}_\qqq (t) \lio_{\text{T}} \pp$.
The propagator for the entangled part is defined as $\bar{G}_\qqq (t)= e^{\left( \lio_{\text{S}}+\lio_{\text{B}}+\qqq \lio_{\text{T}} \qqq \right)  t}$. 
The Nakajima-Zwanzig equation is so far exact to all orders in the tunneling Hamiltonian $H_{\text{T}}$ and the Markovian approximation is not performed so that the time evolution, not only the steady-state, is captured 
exactly via the propagator $\bar{G}_\qqq (t)$. In this Letter, however, only the steady-state is of interest. The steady-state of the reduced density matrix is reached at an infinite time and is per definition $\rho_{\text{red}}(t \rightarrow \infty) = \text{Tr}_{\text{B}}\left\lbrace \rho(t\rightarrow \infty) \right\rbrace  \coloneqq  \rho^\infty_{\text{red}}$. The steady-state of the reduced density matrix allows us to calculate the expectation value of any observable $O$ of the system in the steady-state like the current or the spin: $\expval{O}^\infty = \text{Tr}\left\lbrace \rho^\infty\,O \right\rbrace = \text{Tr}_{\text{S}}\left\lbrace \rho^\infty_{\text{red}}\,O \right\rbrace$. With the help of a Laplace transformation, the convolutive form of the Kernel and the final value theorem, we can simplify \eqref{eq:nakajimazwanzigeq} to 
\begin{equation}\label{eq:steadystate}
\dot{\rho}^{\infty}_{\text{red}} = 0 =\lio \rho^{\infty}_{\text{red}} =\left(  \lio_{\text{S}}+\mathcal{K} \right) \rho^{\infty}_{\text{red}}
\end{equation}
with
\begin{equation}\label{eq:kernelallorders}
\mathcal{K} \rho^{\infty}_{\text{red}} =  \text{Tr}_{\text{B}} \left\lbrace \lio_{\text{T}} \sum_{n=0}^\infty \left(\tilde{\mathcal{G}}_0 \qqq \lio_{\text{T}} \qqq \right)^{2n} \tilde{\mathcal{G}}_0 \lio_{\text{T}} \rho^{\infty}_{\text{red}}\otimes \rho_{\text{B}}\right\rbrace. 
\end{equation}
 $\tilde{\mathcal{G}}_0 = \lim_{\lambda \rightarrow 0^+} \tilde{\mathcal{G}}_0(\lambda)= \lim_{\lambda \rightarrow 0^+} \left[ \lambda - \lio_{\text{S}} - \lio_{\text{B}} \right]^{-1}$ is the free propagator of the system and the bath in Laplace space.
It should be noted that the limit $\lim_{\lambda \rightarrow 0^+}$ should be performed at the very end of the calculation and not in the free propagator alone. In the following, this fact that the limit still has to be performed is marked with $0^+$ in the propagators. 
The form of the propagator $\tilde{\mathcal{G}}_Q(\lambda)$ stems from a Dyson equation. A perturbative approximation is only valid in the so-called weak coupling limit where the tunneling rate is small compared to the temperature $\left(\hbar \Gamma \ll k_{\text{B}} T \right) $.

\subsection{Sequential tunneling}
Considering the full Kernel from \eqref{eq:kernelallorders} to the lowest non-vanishing order, $\mathcal{K}=\mathcal{K}^{(2)}+\mathcal{O}( H_{\text{T}}^4)$, we get the following sequential tunneling Kernel
\begin{equation}\label{eq:secondorder}
\mathcal{K}^{(2)} \rho^{\infty}_{\text{red}}= \text{Tr}_{\text{B}} \left\lbrace \lio_{\text{T}} \frac{1}{0^{+}- \mathcal{L}_{\text{S}}-\mathcal{L}_{\text{B}}} \lio_{\text{T}} \rho^{\infty}_{\text{red}} \otimes \rho_{\text{B}} \right\rbrace.
\end{equation}
This Kernel will be also called second order Kernel due to the appearance of two Louvillians $\lio_{\text{T}}$ and therefore denoted by the superscript $"(2)"$. To simplify the notation of a superoperator $X$, let us introduce the parameter $\alpha$ which is defined by $\comm{X}{\rho}=X\rho-\rho X \coloneqq  X^+\rho-X^-\rho = \sum_\alpha \alpha X^\alpha \rho$. Using this notation for $\lio_{\text{T}}$ yields
\begin{equation}\lio_{\text{T}} X =  - \frac{i}{\hbar} \sum_{\substack{p=\pm \\ \alpha=\pm}} \sum_{\substack{l \sigma k\\n}} p\,c_{l \sigma k}^{p,\alpha} t^{\bar{p}}_{l,n} d_{n}^{\bar{p},\alpha}X.
\end{equation}
In order to simplify the notation, we introduce the index $p$. For example, the superscript $p=+$ indicates the conjugate transpose of the matrix, $d_{n}^{+}\coloneqq  d_{n}^{\dagger}$, or in the case of the tunneling amplitudes, the complex of it, $ t_{l,n}^+\coloneqq t_{l,n}^\ast$.
The superscript $p=-$ indicates an annihilation operator, $d_{n}^{-}\coloneqq  d_{n}$, or a bare tunneling amplitude $ t_{l,n}^{-}\coloneqq t_{l,n}$ with the property of the $p$-index: $\bar{p}\coloneqq-p$. Furthermore, the operators transform to Liouville space superoperators like  $c_{l \sigma k}^{p} \rightarrow c_{l \sigma k}^{p,\alpha}$ where $\alpha$ indicates an operator which acts from the left ($\alpha=+$) or right ($\alpha=-$) side. Applying this on the second order Kernel of \eqref{eq:secondorder} and using the Wick contraction to get the Fermi-function via $\text{Tr} \lbrace c_{l \sigma k}^\dagger c_{l \sigma k} \rho\rbrace=\expval{n}=f^{+}_l(\varepsilon_{l\sigma k})$ we obtain
\begin{equation}
\mathcal{K}^{(2)} \rho^{\infty}_{\text{red}}=  \sum_{\substack{n m p\\l\,\alpha\,\alpha'}} \alpha \alpha'\,\bar \Gamma^{p}_{l,nm} d_{n}^{\bar{p}, \alpha} d_{m}^{p, \alpha'} Y_+^\alpha \left(\chi \right) \rho^{\infty}_{\text{red}}.
\end{equation} 
In this short hand notation of the sequential tunneling Kernel several definitions apply.
Firstly, we use now a different definition of the tunneling rate matrix $\bar\Gamma^{p}_{l,nm} \coloneqq  2 \pi /\hbar \sum_{\sigma l} g_\sigma^0 t_{l,n}^{\ast}t_{l,m}$ with the $p$-index and with the sum over $\sigma$ just running over the constant part of the density of states $g_\sigma^0$ in order to simplify the expressions.
Generally, we define the density of states of the leads as $g_{l\sigma}(\varepsilon)=\rho^0_\sigma  W^2/((\varepsilon-\mu_l)^2+  W^2)$ with the chemical potential $\mu_l$ for the $l$-lead. The Lorentzian cut-off-function is needed for the energy integral evaluation but we argue that it is also justified to introduce it because in real metals the bands are not infinite in terms of energy. The argument of $Y_m^n$ is $\chi=(\Delta E_{m p \alpha'} -p\mu_l )/(k_{\text{B}} T)$ where $T$ is the temperature, $k_{\text{B}}$ is the Boltzmann constant and $\Delta E_{m p \alpha'}$ is the energy difference between the steady-state reduced density matrix and a virtual state where the operator $d_{m}^{p, \alpha'}$ is applied on the former. The energy integration itself is contained in the $Y_m^n$-function with its dimensionless variables $\mu$ and $x$, the dimensionless Fermi-function $f^{(n)} (x)$ and the Lorentzian cut-off-function $ L( \tilde W,x)=\tilde W^2/(x^2+\tilde W^2)$ with the dimensionless high energy band limit $\tilde W=W/(k_{\text{B}} T)$ to ensure the convergence of the integration:
\begin{equation}
Y_m^n\left(\mu \right) \coloneqq -  \frac{i}{2 \pi} \int \diff x \frac{f^{(n)} (x)\,{L}( \tilde  W,x)}{m(x-\mu)+i0^+}.    
\end{equation}
Applying the residuum theorem one gets for $Y_+^n\left(\mu \right)$ \cite{Mantelli2016}, 
\begin{equation}\label{eq:yfunction}
Y_+^n\left(\mu \right) = -  \frac{1}{2} f^n(\mu) - \frac{in}{2 \pi}\left[\text{Re} \Psi^{(0)}\left( \frac{1}{2}+\frac{i \mu}{2 \pi}\right)-C  \right] \nonumber  =-\frac{1}{4} -\frac{in}{2 \pi}\left[\Psi^{(0)}\left(\frac{1}{2}+\frac{i\mu}{2\pi}\right) -C \right]
\end{equation}
with the constant $C=\Psi^{(0)}( \tfrac{1}{2}+\tfrac{ \tilde  W}{2 \pi})$ defined by the renormalized wide band constant $ \tilde  W$. We furthermore introduced the digamma-function
\begin{equation}\label{eq:digamma}
\Psi^{(0)}\left(z\right) \coloneqq  -\sum_{n=0}^{\infty} \frac{1}{n+z}+\sum_{n=1}^{\infty} \log(1+\frac{1}{n}),\;\;z \in \mathbb{C}.
\end{equation}
The constant $C$ always disappears when summing over the $\alpha$-indices. Therefore, we can drop $C$ from the sequential tunneling Kernel calculation. 

\subsection{Cotunneling and pair tunneling}
\label{sec:cotunneling}
If we include the next leading order in the expansion of the Kernel, we get a Kernel which is valid up to the fourth order: $\mathcal{K}=\mathcal{K}^{(2)}+\mathcal{K}^{(4)}+\mathcal{O}\left( H_{\text{T}}^6\right)$.
This regime of tunneling events up to fourth order in $\lio_\text{T}$ is better known as the cotunneling transport regime. In this regime two new processes are included, namely the cotunneling ones and pair tunneling ones. For the fourth order Kernel, we obtain according to \eqref{eq:kernelallorders}: 
\begin{eqnarray}\label{eq:fourthorder}
\mathcal{K}^{(4)} = &\pp& \lio_{\text{T}} \tilde{\mathcal{G}}_0 \qqq \lio_{\text{T}} \qqq \tilde{\mathcal{G}}_0 \qqq \lio_{\text{T}} \qqq \tilde{\mathcal{G}}_0 \lio_{\text{T}} \pp \nonumber \\
= &\pp& \lio_{\text{T}} \tilde{\mathcal{G}}_0 \lio_{\text{T}} \tilde{\mathcal{G}}_0 \lio_{\text{T}} \tilde{\mathcal{G}}_0 \lio_{\text{T}} \pp \nonumber - \pp \lio_{\text{T}} \tilde{\mathcal{G}}_0 \lio_{\text{T}} \pp \tilde{\mathcal{G}}_0  \pp \lio_{\text{T}} \tilde{\mathcal{G}}_0 \lio_{\text{T}} \pp.
\end{eqnarray}
Here we applied $\pp \lio_{\text{T}}^{2n+1} \pp=0$ for $n \in \mathbb{N}$ and the fact that $\mathcal{P}$ respective $\qqq$ commutes with $\tilde{\mathcal{G}}_0$ so that the outermost $\qqq$-operators vanish and the innermost square to $\qqq$. If we let $\mathcal{K}^{(4)}$ act on a density matrix, we get:
\begin{eqnarray}
&~&\mathcal{K}^{(4)}\rho^{\infty}_{\text{red}}= \left[ \mathcal{K}^{(4,D)}+\mathcal{K}^{(4,X)}\right] \rho^{\infty}_{\text{red}} = \sum_{\{\alpha_i\}\{l\} } \sum_{\{n\} \{m\} \{p\}}  \frac{\alpha_1\alpha_4}{k_{\text{B}} T} 
 \bar\Gamma^{p}_{l,nm}\bar\Gamma^{p'}_{l',n'm'}  \nonumber \\ &~& \Bigg[  D^{\alpha_1\alpha_2}_{++}(\square ,\odot,\bullet )d_{n}^{\bar{p}, \alpha_4} d_{n'}^{\bar{p}', \alpha_3} d_{m'}^{p', \alpha_2} d_{m}^{p, \alpha_1}   +X^{\alpha_1\alpha_2}_{++}(\square ,\star,\bullet)d_{n}^{\bar{p}, \alpha_4} d_{n'}^{\bar{p}', \alpha_3}  d_{m}^{p, \alpha_2} d_{m'}^{p', \alpha_1} \Bigg]  \rho^{\infty}_{\text{red}}\label{eq:fourthorderkernel}
\end{eqnarray}
with $\square =(\mu_{j_3}-p\mu_l)/(k_{\text{B}} T),\;\odot =(\mu'_{j_1}-p\mu_{l})/(k_{\text{B}} T),\;\bullet = (\Delta_{j_2}-p\mu_l-p'\mu_{l'})/(k_{\text{B}} T)$ and $\star = (\mu'_{j_1}-p'\mu_{l'})/(k_{\text{B}} T)$.
The subscripts $\left\lbrace j_1,j_2,j_3 \right\rbrace$ of the energy differences $\mu',\Delta$ and $\mu$ indicate that these energies depend on the variables of the first ($\alpha_1,p,m$), two first respective three first $d$-superoperators. More details on the derivation of this result can be found for example in chapter 2 of \cite{Niklas2018}. The $D$-function is defined by
\begin{eqnarray}
\label{eq:D_integral}
&\,& D^{n n'}_{p p'}(\mu,\mu',\Delta) = \nonumber -\frac{i\hbar}{4 \pi^2} \int\displaylimits_{-\infty}^\infty \diff x \int\displaylimits_{-\infty}^\infty\diff x' \frac{f^{\left( n \right)} \left(x \right)}{i 0^{+}+p\left(x-\mu \right)} \frac{1}{i 0^{+}+p x+p'x'-\Delta} \frac{f^{\left( n' \right) }\left(x' \right)}{i 0^{+}+p\left(x-\mu' \right)} \nonumber \\
&\,&= \frac{2 \pi^2 n \left( i\pi+2 C n'\right)}{i\hbar\left(\mu-\mu'\right)}  \left[\Psi^{(0)}\left(\frac{1}{2} +\frac{i\mu}{2\pi}\right) -\Psi^{(0)}\left(\frac{1}{2} +\frac{i\mu'}{2\pi}\right)\right] -\frac{2 \pi n n'}{\hbar} \sum_{k=0}^{\infty}\frac{\Psi^{(0)}\left(1+k +\frac{i\Delta}{2\pi}\right)}{\left( k+\frac{1}{2}+\frac{i\mu}{2\pi}\right)\left( k+\frac{1}{2}+\frac{i\mu'}{2\pi}\right) }
 \end{eqnarray}
and the $X$-function reads
\begin{eqnarray}
\label{eq:X_integral}
&\,&X_{p p'}^{n n'}(\mu,\mu',\Delta)= -\frac{i\hbar}{4 \pi^2} \int\displaylimits_{-\infty}^\infty \diff x \int\displaylimits_{-\infty}^\infty\diff x' \frac{f^{\left( n \right)} \left(x \right)}{i 0^{+}+p\left(x-\mu \right)} \frac{1}{i 0^{+}+p x+p'x'-\Delta} \frac{f^{\left( n' \right) }\left(x' \right)}{i 0^{+}+p'\left(x'-\mu' \right)} \nonumber \\
&\,&  =  -\frac{4 \pi^2}{i\hbar}\frac{ n n'}{\mu+\mu'-\Delta}\Psi^{(0)}\left(\frac{1}{2}+\frac{i\mu}{2\pi}\right) \biggl[\Psi^{(0)}\left(\frac{1}{2}+\frac{i\mu'}{2\pi}\right) -\Psi^{(0)}\left(\frac{1}{2}+\frac{i(\Delta - \mu)}{2\pi}\right) \biggr]    \nonumber \\
&\,&+\frac{2 \pi n n'}{\hbar}\sum_{k=0}^{\infty}\frac{\Psi^{(0)}\left(1+k +\frac{i\Delta}{2\pi}\right)}{\left( k+\frac{1}{2}+\frac{i\mu'}{2\pi}\right)\left( k+\frac{1}{2}+\frac{i(\Delta-\mu)}{2\pi}\right) }. 
\end{eqnarray}
It should be noted that there are many special cases (like $\Delta =0$) where the $D$- and $X$-functions become fully analytical. It can be shown that this superoperator formalism is equivalent to the ansatz in the dissertation of S. Koller \cite{Koller2010_2}. The superoperator formalism, though, treats the problem in the Liouville space and maps into a more compact single-path diagrammatics \cite{Koller2010_1}. Koller's approach is performed in the Hilbert space and maps into a double-path diagrammatics, closer to the real time diagrammatics \cite{Schoeller1996,Koenig1995}. We formulated here also an expression for the imaginary part of the energy integrals \eqref{eq:D_integral} and \eqref{eq:X_integral} of the fourth order Kernel which are needed for the time evolution of fourth order coherences.
\bibliography{bibliography}